\documentclass[preprint]{aastex63}

\usepackage{amsmath}
\usepackage{multirow}

\newcommand\Rone{\textbf{\uppercase\expandafter{\romannumeral1}}\,}
\newcommand\Rtwo{\textbf{\uppercase\expandafter{\romannumeral2}}\,}
\definecolor{forestgreen}{rgb}{0.10, 0.50, 0.10}

\definecolor{forestgreen}{rgb}{0.10, 0.50, 0.10}
\newcommand{\jchen}[1]{#1}

\received{}
\revised{ }
\accepted{ August 28, 2022}
\submitjournal{ApJL}

\shorttitle{FastQSL}
\shortauthors{Zhang et al.}

\begin{document}

\title{FastQSL: A Fast Computation Method for Quasi-separatrix Layers}

\author[0000-0001-6855-5799]{PeiJin Zhang}

\affiliation{Institute of Astronomy and National Astronomical Observatory,\\ Bulgarian Academy of Sciences, Sofia 1784, Bulgaria}
\affiliation{ASTRON, The Netherlands Institute for Radio Astronomy,\\
		Oude Hoogeveensedijk 4, 7991 PD Dwingeloo, The Netherlands}
\affiliation{Astronomy \& Astrophysics Section, Dublin Institute for Advanced Studies, Dublin 2, Ireland. }		
\affiliation{CAS Key Laboratory of Geospace Environment,
	 School of Earth and Space Sciences, \\
	 University of Science and Technology of China,
	  Hefei, Anhui 230026, China}


\correspondingauthor{Jun Chen}
\email{el2718chenjun@nju.edu.cn}
\author[0000-0003-3060-0480]{Jun Chen}
\affiliation{School of Astronomy and Space Science, Nanjing University, Nanjing 210023, China}
\affiliation{CAS Key Laboratory of Geospace Environment,
	 School of Earth and Space Sciences, \\
University of Science and Technology of China,
	  Hefei, Anhui 230026, China}

\author[0000-0003-4618-4979]{Rui Liu}
\affiliation{CAS Key Laboratory of Geospace Environment,
	 School of Earth and Space Sciences, \\
	 University of Science and Technology of China,
	  Hefei, Anhui 230026, China}
\affiliation{CAS Center for the Excellence in Comparative Planetology, 
    \\University of Science and Technology of China,  Hefei, Anhui 230026, China}

\author[0000-0001-6252-5580]{ChuanBing Wang}
\affiliation{CAS Key Laboratory of Geospace Environment,
	 School of Earth and Space Sciences, \\
	 University of Science and Technology of China,
	  Hefei, Anhui 230026, China}
\affiliation{CAS Center for the Excellence in Comparative Planetology, 
    \\University of Science and Technology of China,  Hefei, Anhui 230026, China}


\begin{abstract}
Magnetic reconnection preferentially takes place at the intersection of two separatrices or two quasi-separatrix layers, which can be quantified by the squashing factor $Q$, 
whose calculation is 
computationally expensive due to the need to trace as many 
field lines as possible.
We developed a method (FastQSL) optimized for obtaining 
$Q$  and the twist number in a 3D data cube. FastQSL utilizes the hardware acceleration of the graphic process unit (GPU) and adopts a step-size adaptive scheme for the most computationally intensive part: tracing magnetic field lines. As a result, it achieves a computational efficiency of 
4.53 million  $Q$  values per second.
FastQSL is open source, and user-friendly for data import, export, and visualization.


\end{abstract}



\section{Introduction}
Chromosphere flare ribbons often coincide with the footprints of separatrices or quasi-separatrix layers (QSLs) \citep{priest1995three, demoulin1996three, Demoulin1997}, 
which embed the favorable sites for 3D magnetic reconnection, such as
null points, separators (intersection of two separatrices) and 
 quasi-separators (intersection of two QSLs)
of the magnetic field \citep{Priest2000,Pontin2011}, where a large gradient of magnetic connectivity is present.

The squashing factor $Q$ quantifies the connectivity change of magnetic field lines \citep{Titov2002, Titov2007}, 
$Q$ is defined through a mapping from 
one surface $S_1$, which a magnetic field line threads at $(x_1, y_1)$, to another surface $S_2$, which the field line threads at $(x_2, y_2)$; 
\begin{align}
\underset{1\,2}\Pi: (x_1,y_1) \rightarrow (x_2,y_2).
\label{pi+-}
\end{align}
The Jacobian matrix of differential mapping is expressed as
\begin{align}
\underset{1\,2}{D}=\left(
\begin{array}{cc}
\frac{\partial x_2}{\partial x_1} & \frac{\partial x_2}{\partial y_1} \\
\frac{\partial y_2}{\partial x_1} & \frac{\partial y_2}{\partial y_1} \\
\end{array}
\right)
\equiv
\left(
\begin{array}{cc}
a & b \\
c & d \\
\end{array}
\right).
\label{eq:D+-}
\end{align}

A full description of $Q$ that considers the variance of the covariant metric
tensor on $S_1$ and $S_2$ is defined by Equations (11), (12) and (14) in \citet{Titov2007}.
If $S_1$ and $S_2$ are the two boundaries of a cuboid coordinated 
with the same Cartesian coordinate system, 
the value of $Q$ at $(x_{1},y_{1})$ is
\begin{align}
Q\left(x_1,y_1\right)=\frac{a^2+b^2+c^2+d^2}
{\left|\text{det}\, \underset{1\,2}{D}\right|}. 
\label{eq:Q+-}
\end{align}
$\left|\text{det}\, \underset{1\,2}{D}\right|$ is often replaced by its equivalence $\left|B_{n,1} / B_{n,2} \right|$ for mitigating numerical error,
where
\begin{align}
B_{n,1}=\left.(\Vec{B}\cdot \Vec{n}) \right |_{S_1}
\label{eq:Bn1}
\end{align}
is the component normal to $S_1$ of $(x_1,\, y_1)$, and $\Vec{n}$ is the normal unit vector of ${S_1}$, the form is similar to that of $B_{n,2}$.
Since \cite{Titov2002} proved $Q\left(x_1,y_1\right)\,=\,Q\left(x_2,y_2\right)$,
\cite{Aulanier2005} expanded the definition of $Q$ into 3D space by 
\begin{align}
\vec{B} \cdot \nabla Q = 0,
\label{eq:qline}
\end{align}
i.e. values of $Q$ are invariant along a field line. 
Separatrices are located where $Q=\infty$, 
and
QSLs are located where $Q\gg 2$, the theoretical minimum of Q.

Along with tracing field lines, 
the twist number,
a measure of how many turns two infinitesimally close field lines wind about each other,
can be calculated without much additional effort by (Equation (16) in \cite{berger2006writhe}) 
\begin{align}
\mathcal{T}_w=  \int_L^{}\frac{\nabla\times\vec{B}\cdot\vec{B}}{4\pi B^2}\textrm{d}l,
\label{eq:tw}
\end{align}
where the integral range $L$ is a segment of a magnetic field line. 

In this work, we take advantage of GPU computing,
which is more efficient and economic compared to traditional CPU computing \citep{zwart2020ecological}, to obtain the 3D distribution of $Q$ and $\mathcal{T}_w$ with high efficiency. \cite{feng2013gpu} achieved an acceleration ratio of about 8 times  by re-writing the  model into a GPU compatible form for the magnetohydrodynamics (MHD) simulation of space weather. \cite{Caplan2018GPUAO} accelerated the solar MHD code based on OpenAcc, the GPU version has about 3 times the efficiency of the CPU version, with a comparable cost level of hardware. \cite{tassev2017qsl} have implemented the computation of QSLs with a GPU based on OpenCL, and achieved the efficiency of obtaining a representative 3D QSL map within a few hours.

The rest of this paper is arranged as follows: in Section 2, the algorithm and program structure is presented. In Section 3, we use the extrapolated potential field from a solar active region on 2010 Oct 16 19:00\,UT (AR11112)  and an analytical field from \cite{titov1999TD99} (TD99 model)
to demonstrate the method. In Section 4, we present detailed benchmarks and comparisons for different algorithms and computation architectures. In Section 5, we discuss and summarize the result. 

\section{Method}
FastQSL is developed from the published source code \footnote{ \url{http://staff.ustc.edu.cn/~rliu/qfactor.html} } \citep{Liu2016apj}, hereafter Code2016. 

\subsection{Calculation of \texorpdfstring{$Q$}{Q}} \label{Q-Calculation}
In the computation of $Q$, the essential and most computational consuming step is numerically deriving magnetic field lines by solving the equation:
\begin{align}
    \frac{\textrm{d}\,\Vec{r}\,(s)}{\textrm{d}\,s} = \frac{\Vec{B}}{B},
    \label{eq:rb}
\end{align}
where $s$ is the arc length coordinate of a field line, and
$\vec{r}\,(s)$ is the coordinates function of the field line.

To accurately map the field-line foot-points on a surface, one must solve the field-line equation in high precision.
In Code2016, 
Equation \eqref{eq:rb} is integrated by the classic \texttt{RK4}.
We terminate the integration where it goes outside the data cube, and then move one step back to get the field-line coordinates at the boundary.
%
%

$\underset{1\,2}{D}$ in Equation \eqref{eq:D+-} is derived from 
the changes in the mapped footpoint coordinates with respect to the footpoints of neighboring field lines.
$Q$ on a cross section can be calculated with the method 3 introduced in \cite{Pariat2012} (hereafter Method \Rone), which introduces an auxiliary cross section $S_0\,(x_0,y_0)$ to obtain
\begin{align}
\underset{1\,2}{D}=
\left(
\begin{array}{cc}
\frac{\partial x_2}{\partial x_0} & \frac{\partial x_2}{\partial y_0} \\
\frac{\partial y_2}{\partial x_0} & \frac{\partial y_2}{\partial y_0} \\
\end{array}
\right)
\left(
\begin{array}{cc}
\frac{\partial x_0}{\partial x_1} & \frac{\partial x_0}{\partial y_1} \\
\frac{\partial y_0}{\partial x_1} & \frac{\partial y_0}{\partial y_1} \\
\end{array}
\right),
\label{eq:D+-2}
\end{align}
and
\begin{align}
\left(
\begin{array}{cc}
\frac{\partial x_0}{\partial x_1} & \frac{\partial x_0}{\partial y_1} \\
\frac{\partial y_0}{\partial x_1} & \frac{\partial y_0}{\partial y_1} \\
\end{array}
\right)
=
\left(
\begin{array}{cc}
\frac{\partial x_1}{\partial x_0} & \frac{\partial x_1}{\partial y_0} \\
\frac{\partial y_1}{\partial x_0} & \frac{\partial y_1}{\partial y_0} \\
\end{array}
\right)^{-1}
=
\left.\left(
\begin{array}{rr}
 \frac{\partial y_1}{\partial y_0}   &-\frac{\partial x_1}{\partial y_0} \\
-\frac{\partial y_1}{\partial x_0}   & \frac{\partial x_1}{\partial x_0} \\
\end{array}
\right) \right/ |B_{n,0}/B_{n,1}|,
\label{eq:D+-3}
\end{align}
where $B_{n,0}$ is the component normal to $S_0$, that has a similar form as Equation \eqref{eq:Bn1}. Method \Rone is used in Code2016.

\cite{tassev2017qsl} published their code \texttt{QSL Squasher} 
\footnote{\url{https://bitbucket.org/tassev/qsl_squasher/src/hg/}} that achieved a high efficiency to identify QSLs, 
and \cite{Scott2017} gave a detailed analysis of the code. 
Taking $\Vec{U},\Vec{V}$ as a pair of orthonormal unit vectors on $S_0$,
\cite{Scott2017}
then proposed a method of obtaining $Q$ without the information of neighboring mapping coordinates by solving 
\begin{align}
    \frac{\textrm{d}\{\Vec{r},\, \Vec{U},\, \Vec{V}\}}{\textrm{d}s} = 
    \{\frac{\Vec{B}}{B}, 
    \,\Vec{U} \cdot \nabla \frac{\Vec{B}}{B},
    \,\Vec{V} \cdot \nabla\frac{\Vec{B}}{B}\},
    \label{eq:rb2}
\end{align}
and they proved that
\begin{align}
Q = \frac{ \Tilde{\Vec{U}}_1^2\, \Tilde{ \Vec{V}}_2^2 +
\Tilde{\Vec{U}}_2^2\, \Tilde{ \Vec{V}}_1^2
-
2\,(  \Tilde{\Vec{U}}_1\cdot\Tilde{ \Vec{V}}_1)\,(  \Tilde{\Vec{U}}_2\cdot \Tilde{\Vec{V}}_2)}
{(B_{n,0})^2/ (B_{n,1}\, B_{n,2})}
\label{eq:q_scott}
\end{align}
is equivalent to Equation \eqref{eq:Q+-},
where
\begin{align}
\Tilde{\Vec{U}}_1= \Vec{U}-{\left.
\frac{ \Vec{U}\cdot \Vec{n} }{ \Vec{B}\cdot \Vec{n} } \Vec{B} \right|}_{S_1},
\end{align}
the form is similar for $\Tilde{\Vec{U}}_2,\,\Tilde{\Vec{V}}_1,\,\Tilde{\Vec{V}}_2$.
%
%
In this paper, the method of \cite{Scott2017} based on Equation (\ref{eq:q_scott}) is referred to as Method \Rtwo.

An alternative set of codes for calculating QSLs  is published on \footnote{\url{https://github.com/Kai-E-Yang/QSL}} (here after CodeYang), 
the first version still followed  Method \Rone and was firstly applied in \cite{Yang2015}. 
As of October 2018, CodeYang adopted Method \Rtwo. 

Different selections of $S_1,\,S_2$ will result in different values of $Q$ even
for the same start point on $S_0$ (see the example in Section 4.1 of \citet{Titov2007}).
In Code2016 and FastQSL, $S_1,\,S_2$ by Method \Rone are 
the boundaries where a field line terminates. 
Therefore Code2016 and FastQSL record these boundaries for every field line. 
If one locally rotates $S_1, S_2$ to be perpendicular to the magnetic field line, Equation \eqref{eq:Q+-} and Equation \eqref{eq:q_scott} will give $Q_{\perp}$ \citep{Titov2007}.
$Q_{\perp}$ removes the projection effect on boundaries, therefore quantifies the property of volume QSL more precisely than $Q$ does. The reason that $Q$ is still often used rather than $Q_\perp$ is for its numerical simplicity. \texttt{QSL Squasher} and CodeYang set $S_1,\,S_2$ to always be perpendicular to $\Vec{B}$, therefore providing $Q_\perp$ only. 

Method \Rone requires tracing at least 4 neighboring field lines for the central difference of footpoint coordinates, 
resulting in a numerical difficulty in cases where 4 field lines have different $S_1,\,S_2$, that gives \texttt{NaN} in Code2016.
Method \Rone also has difficulty of accurately applying the formula of  $Q_\perp$ of \citet{Titov2007}, especially on a polarity inversion line (PIL).
Method \Rtwo can give $Q$ and $Q_\perp$ directly without introducing the error of coordinate difference by tracing Equation \eqref{eq:rb2} alone, 
but solving Equation \eqref{eq:rb2} is less efficient than 
Equation \eqref{eq:rb} because of the need of calculating $\nabla \frac{\Vec{B}}{B} $ at every step. 
In addition, since $Q$ changes sharply around a separatrix,
the high-$Q$ positions could be inside cells whose surrounding grids still have low values of $Q$,
then these separatrix segments can not be captured.
In contrast, with Method \Rone,
Code2016 traces field lines at refined grids,
and implement central difference of the mapping coordinates 
in terms of refined grids
for Equations \eqref{eq:D+-2} and \eqref{eq:D+-3}; 
here the grid spacing is denoted as $\delta$ in \citet{Pariat2012}.
Consequently, a separatrix can be captured with
refined girds,
and characterized by an extremely high value of $Q$. 
Briefly speaking, Method \Rone has the advantage of locating the position of thin QSLs especially separatrices (except that $S_1, S_2$ are not exactly same for 4 neighboring field lines), 
Method \Rtwo has the advantage of giving accurate values of $Q$ and $Q_\perp$.
These characteristics are shown in Section \ref{sec:results} (Figure \ref{fig:res2D} and Figure \ref{fig:quadrapole}).
Since $Q$ is mostly used to 
locate 
the position of QSLs and separatrices, 
Method \Rone still has its advantage.
FastQSL provides the option of both methods.

\subsection{Magnetic Field at the Input}
\jchen{
For Code2016 and FastQSL, the input magnetic field is assumed to be in Cartesian grids.
Code2016 requires uniform grid spacings, while FastQSL additionally supports  general stretched (but still rectilinear) grids.
CodeYang and \texttt{QSL Squasher} can run in spherical coordinates, \texttt{QSL Squasher} can run in stretched grids.
}

\jchen{
We assume that the input magnetic field $\Vec{B}$ is known at every 3D Cartesian grid $[x_i,\,y_j,\,z_k]$. 
Then the $\vec{B}$ at $\vec{r}=(x,\,y,\,z)$ in 
a cubic unit cell $[x_i,\,x_{i+1}]\times[y_j,\,y_{j+1}]\times[z_k,\,z_{k+1}]$  is interpolated by
\begin{align}
    \Vec{B}_{\textrm{interp}}(x,\,y,\,z) =& \sum_{m=0}^{1} \sum_{n=0}^{1} \sum_{p=0}^{1} \omega_{x,m}\,
    \omega_{y,n}\, \omega_{z,p}\, \Vec{B}(x_{i+m},\,y_{j+n},\,z_{k+p}), \label{eq:interp}\\
    \omega_{x,0}=&\frac{x_{i+1}-x}{x_{i+1}-x_i},
    \label{eq:w0}\\
    \omega_{x,1}=&1- \omega_{x,0},
    \label{eq:w1}
\end{align}
where $\omega_{y,n},\, \omega_{z,p}$ have similar forms as Equation \eqref{eq:w0} and Equation \eqref{eq:w1}.
For uniform grids, simply flooring $\{x,\,y,\,z\}/\text{spacing}$ is $\{i,\,j,\,k\}$ in Equation \eqref{eq:interp}.
While for stretched grids, we apply a binary search for the determination of $i$, $j$, $k$, which is much more time-consuming than the flooring, and the final performance is reduced to 45\%-80\% (depends on settings at the input) by this determination.
}

\subsection{Tracing Scheme}
\jchen{
Code2016 and CodeYang utilizes the classic \texttt{RK4} to solve Equation \eqref{eq:rb} and Equation \eqref{eq:rb2}. \texttt{QSL Squasher} was updated to version 2.0 from January 2019, then the option of tracing scheme of Cash-Karp method is removed and 
only Euler integration is retained, previous versions are not available online currently. All of them use uniform fixed step-size (here after \texttt{step}).
}

\jchen{
FastQSL updates the tracing scheme with the 3/8-rule \texttt{RK4} 
\citep{kuttaRK4}, which introduces a smaller step error than the classic \texttt{RK4},
and additionally provides \texttt{RKF45} \citep{fehlberg1969low} for further acceleration. \texttt{RKF45} calculates the difference between \texttt{RK4} and \texttt{RK5} of each step, 
\texttt{tol} is the maximum tolerated difference, 
and the unit of \texttt{tol} is the original grid spacing. 
If the difference is larger than \texttt{tol}, 
that is a failed-trial step,
then the step-size is adjusted to a smaller value and repeat the tracing step from the same point.
If the difference is smaller than \texttt{tol}, 
\texttt{RKF45} accepts this tracing step and then adjust the step-size to a larger value 
according to the last difference and \texttt{tol}.
A smaller value of \texttt{tol} will result in a more precise output, but takes more computational resources.
}

\jchen{
If grids are stretched, we adopt 
a self-adaptive fashion of  \texttt{step} and \texttt{tol} in cells of different shapes for a better performance. The scaling of \texttt{step} and \texttt{tol} in a cell 
are:
\begin{align} 
\text{scaling}&=
1 \left/\sqrt{
\left(\frac{B_x / B}{x_{i+1}-x_i}\right)^2+
\left(\frac{B_y / B}{y_{j+1}-y_j}\right)^2+
\left(\frac{B_z / B}{z_{k+1}-z_k}\right)^2}\right.,\\
\texttt{tol}_\text{cell}&=\texttt{tol}\times \text{scaling},\\
\texttt{step}_\text{cell}&=\texttt{step}\times \text{scaling}.
\end{align}
\texttt{tol} and \texttt{step} here are dimensionless, $\texttt{tol}_\text{cell}$ and $\texttt{step}_\text{cell}$ that actually applied in the cell have the same unit as $x_i,\,y_j,\,z_k$.
}

\subsection{Code and Algorithm}

\jchen{Code2016 and FastQSL also provide $\mathcal{T}_w$ and  field-line length,
which are also extended to 3D by remaining constant along a field line, like Equation \eqref{eq:qline}. Different from Q-calculation, 
a low-level of (even without) refinement of $\mathcal{T}_w$ is good enough for data 
analysis.
FastQSL additionally provides the coordinates of ending points of field lines, which had been used to locating flaring ribbons in an MHD simulation \citep{Jiang2021}, 
and can help the calculation of slip-squashing factors \citep{Titov2009} if the flows at the boundaries are known.
}

FastQSL provides 2 sets of code. 
The first set is directly developed from Code2016
and still runs on IDL \jchen{(the reliance of SolarSoftWare of Code2016 is removed)}+{Fortran}\footnote{\url{https://fortran-lang.org/}}
This set is optimized in many details. For example, it can be compiled by both \texttt{gfortran} and \texttt{ifort} 
while Code2016 is designed only for \texttt{ifort}.

\begin{figure}
    \centering
    \includegraphics[width=0.99\linewidth]{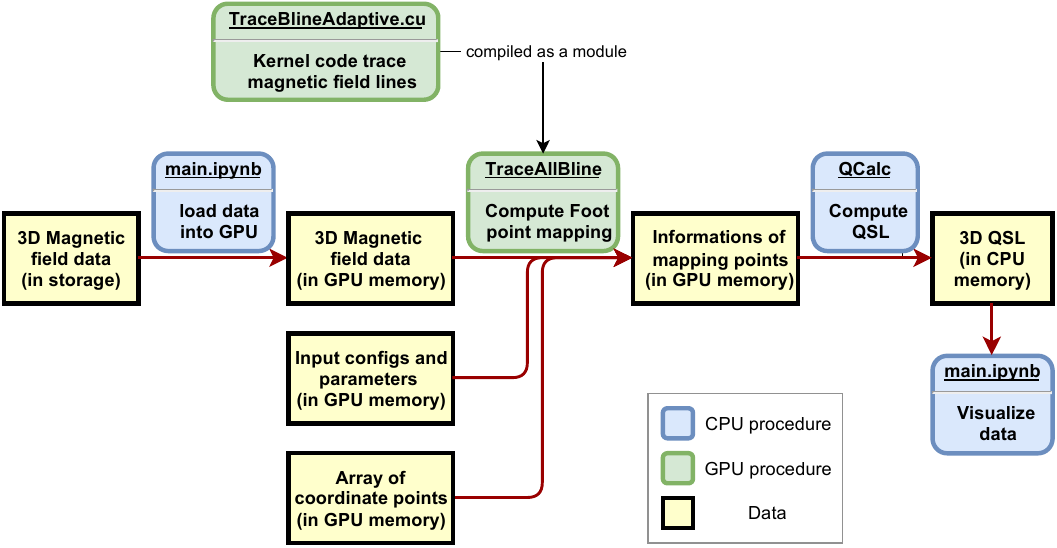}
    \caption{The flowchart of GPU program with Method \Rone in FastQSL. 
    For Method \Rtwo, the words in the upper part of the green blocks should be ``TraceBlineScott.cu" and ``TraceBlineScott", and the block of ``QCalc" is unnecessary.
    }
    \label{fig:chart}
\end{figure}

The 2nd set is accelerated by NVIDIA GPU.
The flowchart of the GPU program with Method \Rone is shown in Figure \ref{fig:chart}, 
the differences from Method \Rtwo is described in the caption of Figure \ref{fig:chart}.
The program is based on \href{http://python.org/}{Python}  and \href{https://docs.nvidia.com/cuda/cuda-c-programming-guide/index.html}{CUDA/C}.  The major input of this program is the data-cube of the 3D magnetic field. 
The data is loaded with \href{https://docs.scipy.org/doc/scipy/reference/io.html}{Scipy.io}, which is capable of reading various file format of data (e.g. .sav .mat and unformatted). For the preparation of GPU computing, the field data, the parameters and the coordinate array of points to be calculated are then transferred to GPU-memory. 
The most computational-power consuming part is to trace magnetic field lines, for which the \texttt{RKF45} solver for is implemented in CUDA/C and compiled as a callable module \texttt{TraceAllBline} (as the green blocks shown in Figure \ref{fig:chart}) by {Cupy}\footnote{\url{https://cupy.dev/}} The compiled module calculates magnetic field lines and derives the foot-point mapping. Then the foot-point mapping results are transferred back to host-memory for the calculation of $Q$.
Also, with the magnetic field line computed in \texttt{TraceAllBline}, $\mathcal{T}_w$ can be simply derived with Equation \eqref{eq:tw}.
After the calculation, $Q$ and $\mathcal{T}_w$ are visualized with {matplotlib} \footnote{\url{https://matplotlib.org/}} (for 2D) and {pyvista}\footnote{\url{https://www.pyvista.org/}} (for 3D).\footnote{Since the server of \url{http://staff.ustc.edu.cn/} will stop after September 2022, the first set will be updated at \url{http://github.com/el2718/FastQSL}. For the second set, the source code, demo example, and document of the method implementation is online available at \url{https://github.com/peijin94/FastQSL}.}

\section{Results} \label{sec:results}
We apply FastQSL to 3 common scenarios of magnetic field analyzing: 
(1) a potential field extrapolated from a solar active region; 
\jchen{(2) an analytical quadrapole field;}
(3) a flux rope from a TD99 model.  
These magnetic fields are used to demonstrated and benchmark FastQSL.

\begin{figure}[h]
    \centering
    \includegraphics[width=0.85\linewidth]{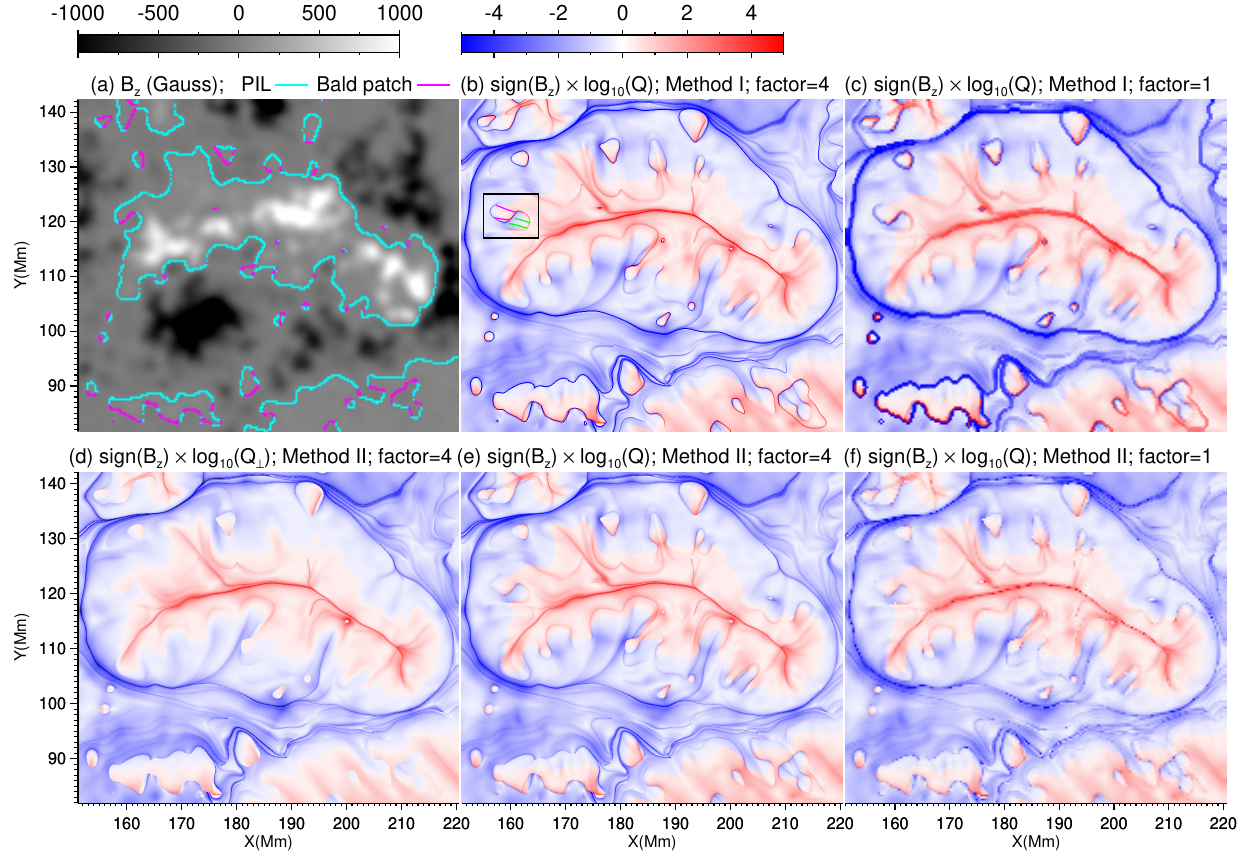}
    \caption{
    Magnetic configuration of NOAA AR 11112 on 2010 October 16 at 19:00 UT. 
    (a) PILs (cyan) and the foot prints of 
    bald patch separatrix (purple) superimposed on the magnetogram. 
    (b-f) Map of QSL at the photosphere, 
    the method and the factor of refinement of grids 
    (refined grid spacing $= 1/\text{factor}\,\times$ original grid spacing)
    are denoted at the image title.
    The coordinates is same as that in \cite{chen2020extreme}. 
    The green and the purple field lines present a double check of the existence of 
    a bald patch in the black box in panel (b).
    }
    \label{fig:res2D}
\end{figure}

\begin{figure}[ht]
    \centering
    \includegraphics[width=0.8\linewidth]{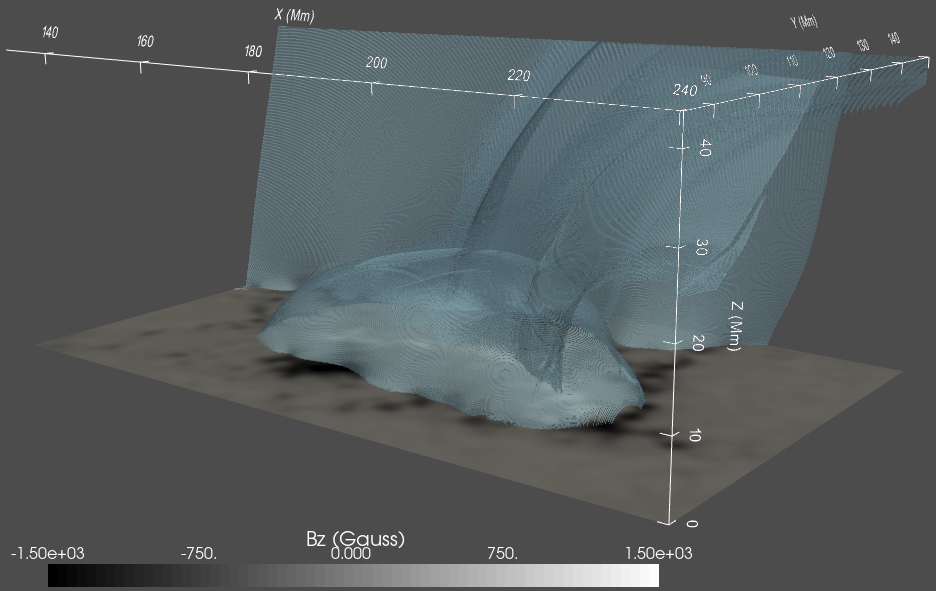}
    \caption{3D topology of the active region.
    The semi-transparent surface is the contour of $Q = 5000$. The computation regime is $900\times540\times360$ pixels, total computation time is 107s.
    The coordinate is same as that in \cite{chen2020extreme}.
    }
    \label{fig:POT3d}
\end{figure}

\subsection{An Extrapolated Potential Field}
The first test is done with the potential field extrapolated from an active region (NOAA AR 11112) on 2010 October 16 at 19:00 UT, which is the same field for 
analyzing ``dome-plate QSL" 
 in \citet{chen2020extreme}. Strictly speaking, it should be ``dome-plate separatrices'' with singular $Q$ since there is a null point \citep{Titov2011,Scott2021}.
As shown by Figure \ref{fig:res2D}(d), $Q_\perp$ can not show any footprints of bald
patch separatrix \citep{Titov&al1993,titov1999TD99}.
For example, in the region bounded by the black box in Figure \ref{fig:res2D}(b), 
there is a footprint of bald patch separatrix on a PIL between 
the green and the purple field lines (the neighboring field lines at two sides of a PIL that is the footprints of a bald patch separatrix should depart with each other), which also satisfies 
$(B_x, B_y, 0) \cdot \nabla B_z|_\text{PIL} > 0$ (the purple lines in Figure \ref{fig:res2D}(a)),
while such topology is missing in Figure \ref{fig:res2D}(d).
As in the discussion above, the footprints of QSLs in Figure \ref{fig:res2D}(d)(e) are much thinner than that in Figure \ref{fig:res2D}(b), even appear discontinuous at the thinnest points. 
Figure \ref{fig:res2D}(c)(f) are calculated with original grids. 
Figure \ref{fig:res2D}(c) keeps the continuity of footprints, 
while Figure \ref{fig:res2D}(f) shows much more discontinuities
than Figure \ref{fig:res2D}(e), 
indicating that the method \Rtwo requires a higher level of refinement for displaying continuous separatrices.

With the significant improvement in efficiency, the 3D distribution of Q can be obtained with ease within the timescale of minutes.  
Figure (\ref{fig:POT3d}) shows the 3D magnetic topologies above the magnetogram.
As shown in Figure \ref{fig:POT3d}, 
the dome structure is well-represented as Figure 4 of \citet{chen2020extreme}. 

\subsection{An Analytical Quadrapole Field}
\begin{figure}[ht]
    \centering
    \includegraphics[width=\linewidth]{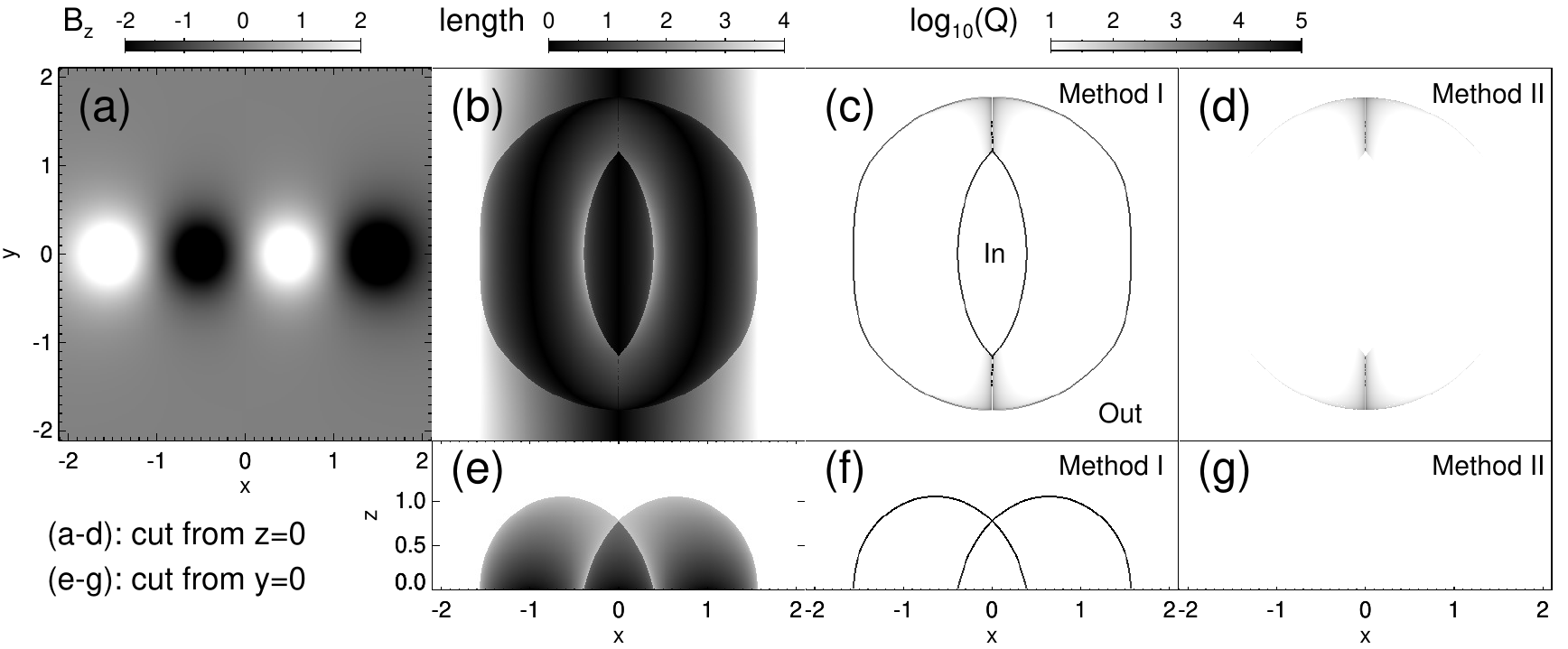}
\caption{Magnetic structures of the quadrapole field. (a) Magnetogram. (b)(e) Field-line length map. (c)(f) Map of $\text{log}_{10}(Q)$ from Method \Rone. (d)(g) Map of $\text{log}_{10}(Q)$ from Method \Rtwo.}
\label{fig:quadrapole}
\end{figure}

\jchen{
In the first case, $Q$ decays exponentially away from the null point \citep{Pontin2016}, there are still some remnant of the high-Q region around separatrices can be captured by Method \Rtwo.
An analytical quadrapole field can  clearly present the shortage of Method \Rtwo on capturing separatrices.
The field is
\begin{align} 
\vec{B}_\text{quadrapole}(\vec{r})&=  \sum_{i=1}^{4}\, q_i\, \frac{\vec{r}-\vec{r}_i}{|\vec{r}-\vec{r}_i|^3},
\label{eq:quadrapole}
\end{align}
where $\vec{r}_1=(-1.5,\,0,\,-0.5)$, $\vec{r}_2=(-0.5,\,0,\,-0.5)$, $\vec{r}_3=(0.5,\,0,\,-0.5)$, $\vec{r}_4=(1.5,\,0,\,-0.5)$ are the locations of 4 magnetic charges, and $\{q_1,\,q_2,\,q_3,\,q_4\}=\{1,\,-1,\,1,\,-1\}$ are the strengths of the magnetic charges. This analytical field is uniformly discretised for FastQSL, and the grid spacing is 0.02. 
There are 4 sun spots on the photosphere (Figure \ref{fig:quadrapole}(a)), 
and the length of field line can sharply jump at some places (Figure \ref{fig:quadrapole}(b)(e)), where must be separatrices. Method \Rone can fully capture all these separatrices (Figure \ref{fig:quadrapole}(c)(f)). 
While Figure \ref{fig:quadrapole}(g) that from Method \Rtwo present a blank because all calculated $Q$ are below 10 in the region of Figure \ref{fig:quadrapole}(g), the case is same if we plot $\text{log}_{10}(Q_\perp)$.
In Figure \ref{fig:quadrapole}(c), in the area closed by the inner separatrices  that labeled with ``in" or in the area outside of the outer separatrices that labeled with ``out",
considering the symmetry of the magnetic field, 
the field-line mapping \eqref{pi+-} should be $x_2=-x_1$, $y_2=y_1$.
According to Equation \eqref{eq:Q+-}, the values of $Q$ in the area ``in" and ``out" should identically be 2.
The real distribution of $Q$ around these separatrices should like the Dirac Delta funtion.
As the discussions of Section \ref{Q-Calculation}, since most refined girds are not on the zero-thickness separatrices, Method \Rtwo can not capture these separatrices (Figure \ref{fig:quadrapole}(d)(g)).
}

\begin{figure}[ht]
    \centering
    \includegraphics[width=0.5\linewidth]{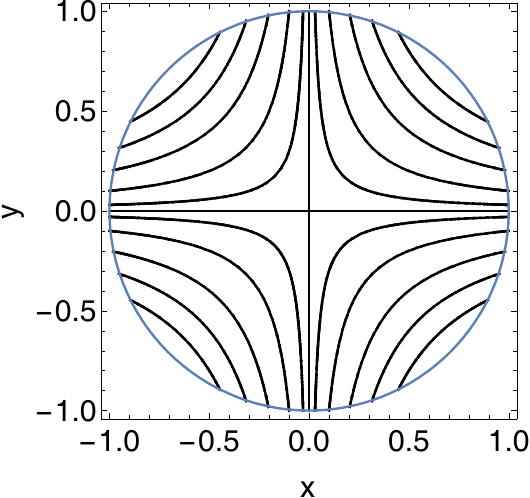}
\caption{Magnetic field lines of $\vec{B}=(x,\,-y,\,0)$. Black curves are field lines, the blue circle is the mapping surface of $Q$-calculation.}
\label{fig:null2d}
\end{figure}
\jchen{
Another case that $\vec{B}=(x,\,-y,\,0)$ has a similar distribution of $Q$. 
We set $x^2+y^2=1$ as 
$S_1(\theta,\,z)$, $S_2(\theta,\,z)$ of $Q$-calculation (Figure \ref{fig:null2d}), 
where $\theta$ is the radian measure.
For $0<\theta<\pi/2$  (the case is similar for $\pi/2<\theta<2\,\pi$), according to the symmetry of the magnetic field, the mappings are $\theta_2=\pi/\,2-\theta_1$, $z_2=z_1$,
and all values of Q are 2 with Equation \eqref{eq:Q+-}. 
Separatrices only appear at $\theta=0,\,\pi/\,2,\,\pi,\,3\,\pi/\,2$.
} 

\subsection{A TD99 model}
\begin{figure}[ht]
    \centering
    \includegraphics[width=0.98\linewidth]{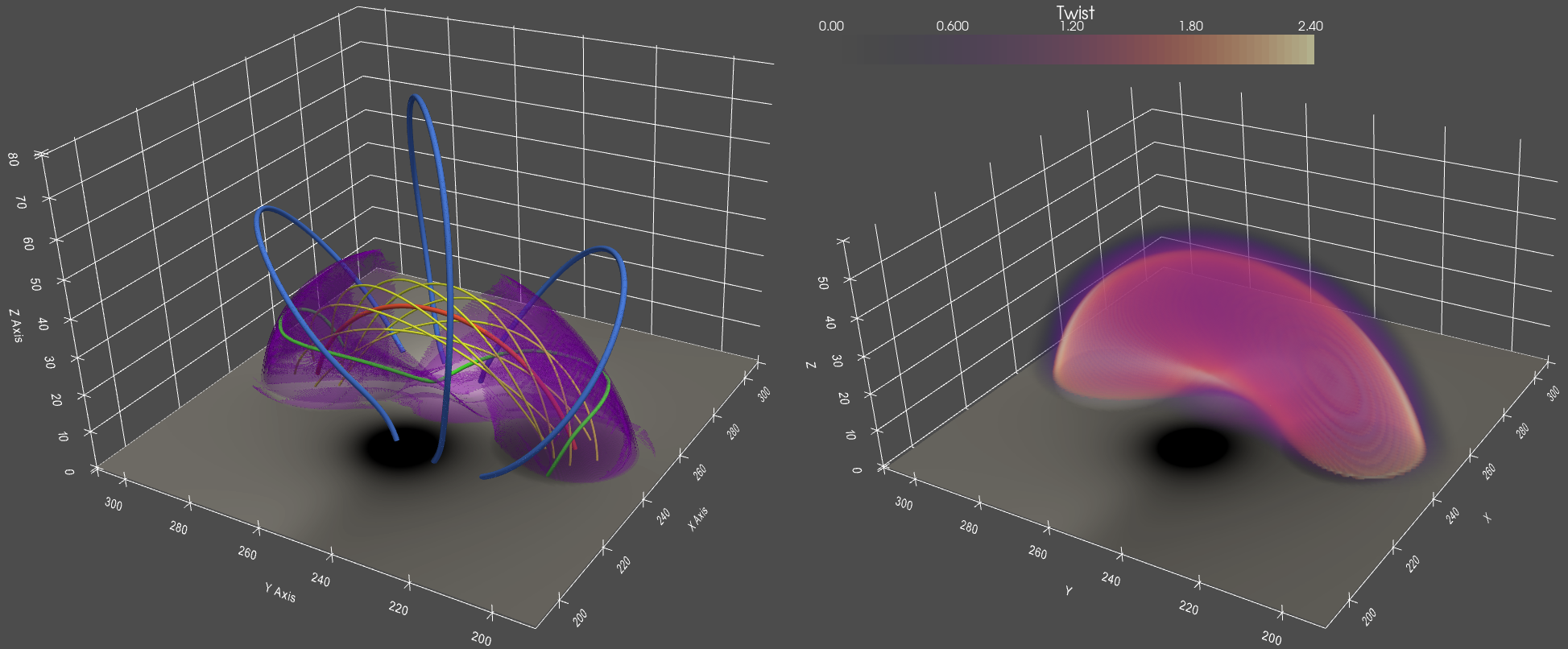}
    \caption{3D structures of the TD99 model. Left: the purple semi-transparent surface is the contour of $Q = 5000$, different types of magnetic field line are rendered in different colors. 
    Right: the distribution of twist number.
    }
    \label{fig:TD3D}
\end{figure}
A TD99 model at the setting of
 $R=110~\rm Mm$, $d=34~\rm Mm$, $L=55~\rm Mm$, 
 $a=49.4~\rm Mm$, 
$I=4\times 10^{12}~A$, 
$I_0=1.66\times 10^{12}~A$, $q= 10^{14}~T~\rm m^2$ is also tested.
\jchen{
This analytical field is uniformly discretised, the grid spacing is 3.04~Mm.}
This setting represents a hyperbolic flux tube (HFT) \citep{Titov2002} topology around the flux rope, 
3D structure of the HFT and the 3D distribution of twist number are presented by Figure \ref{fig:TD3D}.

\section{Benchmark}
\begin{figure}[ht]
    \centering
    \includegraphics[width=\linewidth]{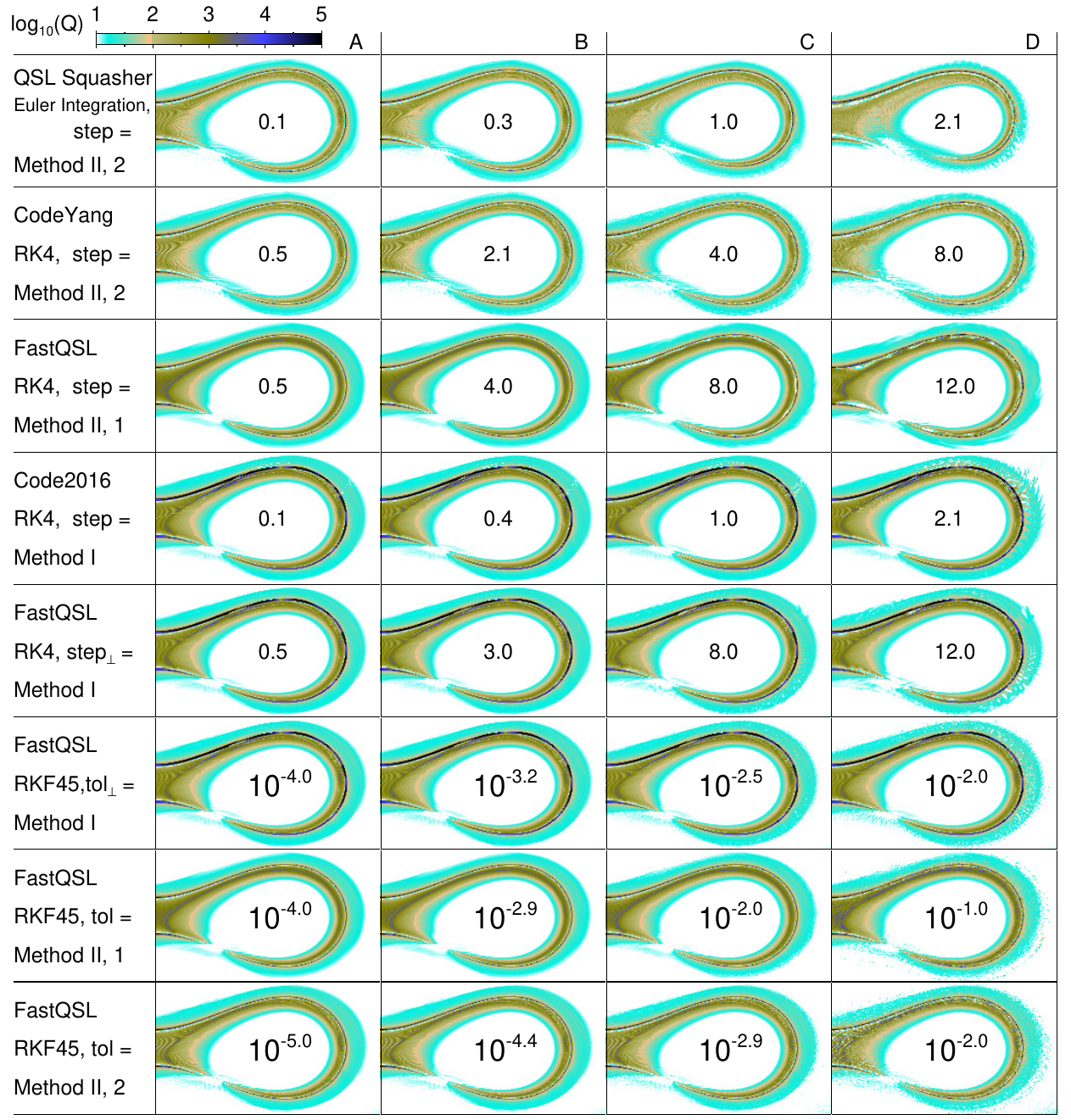}
    \caption{The quality of Q-map with different codes, parameters and methods. 
    The rows of \texttt{QSL Squasher} and CodeYang show $\text{log}_{10}(Q_{\perp})$, other rows show $\text{log}_{10}(Q)$.
    Code and method are marked at the left of image, the value of \texttt{step} or \texttt{tol} is in each panel.
    This test is done with the TD99 model, cut from a cross section at height of 30.4~Mm, 
    the unit of \texttt{step} and \texttt{tol} is the original grid spacing of magnetic field (corresponding to 3.04~Mm).
    Column B presents images with a marginal value that without any indistinct area, comparing to column A. 
    The numbers following method \Rtwo indicate the 1st or the 2nd way of calculating $\nabla \frac{\Vec{B}}{B}$.
    }
\label{fig:quality}
\end{figure}


A benchmark is presented for comparing the efficiency of FastQSL with published codes
(i.e. \texttt{QSL Squasher}, CodeYang and Code2016). The quality of resultant images and the time-consumed depend on the choices of \texttt{step} of \texttt{RK4} or \texttt{tol} of \texttt{RKF45} and method.
Images with different choices of parameters are shown in Figure \ref{fig:quality}.
There is a marginal value of \texttt{step} or $\texttt{tol}$
for the quality of resultant image.
\jchen{For a specific row, comparing horizontally,} comparing with those resultant images with any lower values,
the image with the marginal value should not show any recognizable difference.
And above the marginal value, the difference is recognizable. 
 \jchen{In Figure  \ref{fig:quality}, columns are labeled with ``A, B, C, D'' to mark an image 
 that is ground truth, marginal ground truth, distinguishable and unlikeness.} 
The marginal value depends on the smoothness of the field.
For example, \jchen{comparing to column A and B, }most area in column C are satisfying, but some areas are not, 
one should check the quality of image to make decision. 
 \jchen{
 The image quality is not very sensitive to the value of \texttt{step} or \texttt{tol}, 
 which the efficiency is sensitive to.
 For some analysis that do not require a high quality, even column D can provide an acceptable result, and can achieve a very fast performance.  }

For Method \Rone,
if we fix the tracing parameter \texttt{step} or \texttt{tol},
we find the angle $\theta=\arcsin(|B_{n,0} / B|)$ between the magnetic field line and the cross-section $S_0$ will affect image quality, with a smaller $\theta$ giving poorer results.
To mitigate this effect,
our empirical formulas are:
\begin{align}
\texttt{step}|_{S_0}  &=  \text{max}([\texttt{step}_{\perp} \times | B_{n,0} / B |,\,  \texttt{step}_\text{min}]),  \label{eq:step}\\%
\texttt{tol}|_{S_0} &= \texttt{tol}_{\perp} \times | B_{n,0} / B |^{1.5}, \label{eq:tol}
\end{align}
where $\texttt{step}_{\perp}$, $\texttt{step}_\text{min}$ and $\texttt{tol}_{\perp}$ are constants, \text{max()} is the operation of taking the maximum value. 
When field lines are tangent to $S_0$, $B_{n,0} \to 0$, 
introducing spurious high-$Q$ structures.
In order to avoid this artifact, when $|B_{n,0} / B| < 0.05$, 
we locally rotate $S_0$ so that it is perpendicular to the field line,
and traces 4 neighboring field lines at the new cross-section, and then calculate $Q$.
Therefore, the \texttt{step}  or the \texttt{tol} at the input of FastQSL is for the perpendicular case,
then adjusted by Equation \eqref{eq:step} or \eqref{eq:tol} for every field line. 
Since Code2016 fixes the \texttt{step}, it needs a small marginal \texttt{step} of 0.4 in Figure \ref{fig:quality}.
For Method \Rtwo, FastQSL always sets $S_0$ to the perpendicular case, this adjustment is not applied.

For Method \Rtwo, there can be two strategies to calculate $\nabla  \frac{\Vec{B}}{B}$.
One prepare a 3D array of $\nabla  \frac{\Vec{B}}{B} $ at the beginning by a second order \jchen{finite difference, for example, the formula for uniform grids is}
\begin{align}    
\frac{\partial\, \Vec{B}/B}{\partial\, x}(x_i,\,y_j,\,z_k)
= \frac{1}{(x_{i+1}-x_{i-1})}\left[\frac{\Vec{B}(x_{i+1},\,y_j,\,z_k)} { B(x_{i+1},\,y_j,\,z_k)}-
   \frac{\Vec{B}(x_{i-1},\,y_j,\,z_k)} { B(x_{i-1},\,y_j,\,z_k)}\right],
\label{eq:grad_B}
\end{align}
the form is similar for $\frac{\partial\, \Vec{B}/B}{\partial\, y}(x_i,\,y_j,\,z_k),\, \frac{\partial\, \Vec{B}/B}{\partial\, z}(x_i,\,y_j,\,z_k)$.
Then does an interpolation that is similar to Equation \eqref{eq:interp}.
The other is to interpolate $ \frac{\Vec{B}}{B}$ at the neighboring points of $\vec{r}$ like
$\vec{r}\pm(0.001,0,0), \vec{r}\pm(0,0.001,0), \vec{r}\pm(0,0,0.001)$,
then $\nabla  \frac{\Vec{B}}{B}$ is given by central differences.
the first way gives a smoother distribution than the 2nd.
As shown by the bottom 2 rows of Figure \ref{fig:quality},
the marginal \texttt{tol} by the first way is $10^{-2.9}$ 
while that by the second way is $10^{-4.4}$, 
which means the image quality by the first way is better with the same \texttt{tol}.
The second way needs to calculate additional 6 values of $\frac{\Vec{B}}{B}$,
therefore it is slower than the first way.
But the first way asks additional storage for the 3D array of
$\nabla \frac{\Vec{B}}{B}$
(3 times as the array of $\vec{B}$ occupied) while the second does not.
CodeYang applies the second way.
As the gradient of Equation \eqref{eq:interp} can be analytically given in every cell, 
\texttt{QSL Squasher} applies a skill that is mathematically similar as the second way, 
but the calculation of one value of $\nabla \frac{\Vec{B}}{B}$ consumes the similar duration 
as the first way.
%
%
The first set of FastQSL uses the first way, while the second set applies the second way due to the limitation of GPU memory. 
Comparing with Method \Rone (the row of Code2016 of Figure \ref{fig:quality}),
Method \Rtwo (the top 2 rows of Figure \ref{fig:quality}) allows a much larger \texttt{step}.

Providing a completely impartial benchmark to all codes and both methods is extremely difficult, we just show the benchmark at the marginal values  \jchen{(column B of Figure \ref{fig:quality})}, to make a sense of time-consumed.
By using the TD99 model as an input, 
and calculating for the same region (the 3D region of Figure \ref{fig:TD3D}) and at the same resolution, 
the efficiency is measured by the count of calculated values of $Q$ per unit time.

The original CodeYang can not accept a \texttt{step} $> 1$, because it extends only 1 ghost layer to boundaries. We modified the code to have 10 ghost layers for the benchmark.

\texttt{QSL Squasher} allows adaptive mesh refinement, the possible locations of QSL are quickly identified by Field-line Length Edge (FLEDGE) map, then only calculate $Q_\perp$ at these locations, 
\cite{tassev2017qsl} claimed an order-of-magnitude speed-up with adaptive refinements.
We firstly do not apply adaptive refinements, 
directly set the grid resolution for $Q_\perp$ that is same as other panels of Figure \ref{fig:quality}, 
then achieve the marginal \texttt{step} of 0.3, and use this setting for the benchmark (Tabel \ref{tb:bench}). 
A larger step can shrink the resulting QSLs more significantly (Figure \ref{fig:quality}), 
even slightly affects at the \texttt{step} of 0.3. 
\jchen{We infer this artifact comes from the relatively large step error of Euler integration, because our codes will also have such shrinking if we change the tracing scheme to Euler integration}.
We also try the adaptive refinements, 
at the proper choices of the threshold of length jump for a refinement and the maximum times of refinements for giving a satisfying image that likes column B of Figure \ref{fig:quality},
the performance by GPU is 25 kQ/s, which is surprisingly lower than 189 kQ/s in Tabel \ref{tb:bench}, which is without the adaptive refinements. 
We guess that the gradient of field-line length in the thick QSL could be small still,
then the adaptive refinements can not improve the performance.

As shown by Table \ref{tb:bench},
Code2016 is faster than \texttt{QSL Squasher} and CodeYang.
\texttt{QSL Squasher} is slowed down by Euler integration, double-precision floating-point format, adaptability for \jchen{stretched} girds, the way of calculating $\nabla \frac{\Vec{B}}{B}$, and other potential aspects.
CodeYang is slowed down by double-precision floating-point format,
the way of calculating $\nabla \frac{\Vec{B}}{B}$ and other potential aspects. 
Comparing to Code2016, FastQSL achieves a significant speed-up.
For the first set of FastQSL, 
compiled by \texttt{ifort} is sightly faster than by \texttt{gfortran}.
Traced by \texttt{RKF45} is sightly faster than by \texttt{RK4}, 
but this comparison may be not true for all kinds of magnetic field. 
For a highly twisted field, many failed-trial steps could happen in that by \texttt{RKF45}, 
then by \texttt{RK4} could be even faster. 
If traced by \texttt{RK4}, 
the efficiency by Method \Rtwo is sightly faster than that by Method \Rone,
the comparison is reversed if traced by \texttt{RKF45}.
For the second set of FastQSL that is optimized with GPU, 
it achieves the best efficiency by Method \Rone.
If $Q$ is calculated  by Method \Rtwo, since the second set uses the second way to calculate $\nabla \frac{\Vec{B}}{B}$, it is even slower than the first set.

\begin{table}
\centering
\begin{tabular}{|c|c|c|c|c|c|c|}
\hline
Code & Processor & Compiler  & Method & Tracing scheme & Parameter & Performance\\
\hline
\multirow{2}{*}{\texttt{QSL Squasher}}
 & CPU & \multirow{2}{*}{OpenCL}   
 & \multirow{3}{*}{\Rtwo,\, 2} &  \multirow{2}{*}{Euler integration} &  \multirow{2}{*}{\texttt{step} = 0.3} &  29 kQ/s \\
 \cline{2-2}\cline{7-7}
\multirow{2}{*}{}
 & GPU & \multirow{2}{*}{}  
 & \multirow{3}{*}{} & \multirow{2}{*}{} & \multirow{2}{*}{} & 189 kQ/s \\
\cline{1-3} \cline{5-7}
CodeYang
& \multirow{7}{*}{CPU} & \texttt{gfortran} &  \multirow{3}{*}{} & \multirow{2}{*}{classic \texttt{RK4}} & \texttt{step} = 2.1 & 27 kQ/s \\
\cline{1-1} \cline{3-4}  \cline{6-7}
Code2016
& \multirow{7}{*}{} & \texttt{ifort} & \multirow{3}{*}{\Rone} & \multirow{2}{*}{} & \texttt{step} = 0.4 & 191 kQ/s\\
\cline{1-1} \cline{3-3} \cline{5-7}
\multirow{7}{*}{FastQSL}
 & \multirow{7}{*}{} & \texttt{gfortran} & \multirow{3}{*}{} 
 & \multirow{3}{*}{3/8-rule \texttt{RK4}}
 & \multirow{2}{*}{$ \texttt{step}_\perp = 3.0 $}  &  749 kQ/s\\
\cline{3-3} \cline{7-7}
\multirow{7}{*}{} & \multirow{6}{*}{} & \multirow{4}{*}{\texttt{ifort}}  & \multirow{3}{*}{} & \multirow{3}{*}{} & \multirow{2}{*}{} &  854 kQ/s\\
\cline{4-4} \cline{6-7}
\multirow{7}{*}{} & \multirow{6}{*}{} & \multirow{4}{*}{} & \multirow{2}{*}{\Rtwo, 1} & \multirow{3}{*}{} &  $ \texttt{step} = 4.0 $ &    907 kQ/s\\
\cline{5-7}
\multirow{7}{*}{} & \multirow{7}{*}{} &  \multirow{4}{*}{}  & \multirow{2}{*}{} & \multirow{4}{*}{\texttt{RKF45}} &
$\texttt{tol} = 10^{-2.9}$ & 1.11 MQ/s\\
\cline{4-4} \cline{6-7}
\multirow{7}{*}{} & \multirow{7}{*}{} &  \multirow{3}{*}{} & \multirow{2}{*}{\Rone} & \multirow{4}{*}{} & 
 \multirow{2}{*}{$\texttt{tol}_\perp = 10^{-3.2}$} & 1.43 MQ/s \\
\cline{2-3} \cline{7-7}
\multirow{7}{*}{} & \multirow{2}{*}{GPU} & \multirow{2}{*}{CUDA/C} & \multirow{2}{*}{} &\multirow{4}{*}{}  & \multirow{2}{*}{} &  4.53 MQ/s\\
\cline{4-4} \cline{6-7}
\multirow{7}{*}{} & \multirow{2}{*}{} & \multirow{2}{*}{} & \Rtwo,\, 2
& \multirow{4}{*}{} & $\texttt{tol} = 10^{-4.4}$ & 1.13 MQ/s\\
\hline
\end{tabular}
\caption{The computation efficiency of different methods, measured as the capability of calculating $Q$ per second. 
In this test, the CPU  is Intel core i9 10900K, the GPU is RTX 3070 OC.
The version of \texttt{gfortran} is 9.4.0,
the version of \texttt{ifort} is 2021.6.0.
The columns of ``Processor", ``Compiler" and ``Performance" are only for the computation of QSL, others like IO, preprocessing, visualization are not involved. 
The numbers following \Rtwo indicate the 1st or the 2nd way of calculating 
$\nabla \frac{\Vec{B}}{B}$.
The column of ``Parameter" shows the marginal value of a satisfied image in column B of Figure \ref{fig:quality}. 
\texttt{QSL Squasher} is tested with a smaller data cube of the same TD99 due to its high requirement of memory.
}
\label{tb:bench}
\end{table}

\section{Discussion and Summary}

Comparing with Code2016, the computational efficiency increased by 24 times in this work. The increase in computational efficiency majorly comes from two aspects: new computing architecture and the improvement of algorithm. 
In the computational tasks of data inspection like QSL identification and quantification, GPU acceleration can improve the efficiency of interactive inspection and help to discover new features in data. 
In the computational tasks of simulation \citep{feng2013gpu}, the massive parallel computation with GPU can help explore more parameter space with given time and resource budget. Also, GPU computing is more environmentally friendly by reducing carbon emission \citep{stevens2020imperative,2020NatAsGPU}. 

To summarize, we developed a reliable method of calculating the squashing factor $Q$ and twist number 
in data cubes of magnetic fields on \jchen{Cartesian} grids. This method can achieve unprecedented computation efficiency, with which one can obtain maps  of $Q$ and twist number within a few seconds for 2D input and a few minutes for 3D input. The high efficiency can benefit the analysis of magnetic topology, especially for the analysis of MHD simulations, which may require the computation of 3D $Q$ and twist number in a time series.

\section{Acknowledge}
J.C. thanks the constructive discussion with Guo, Yang, and acknowledges the support by the China Scholarship Council (No.201706340140). 
The research was supported by 
the National Natural Science Foundation of China (42188101, 41974199, 41574167, 41774150, and 11925302), 
the B-type Strategic Priority Program of the Chinese Academy of Sciences (XDB41000000), and STELLAR project (952439).

\bibliography{cite}

\clearpage

\end{document}